\documentclass[times, twoside]{StyleArXiv}
\usepackage{booktabs}
\usepackage{multicol,multirow}
\usepackage{orcidlink}
\usepackage{subcaption}
\usepackage{cleveref}
\crefname{figure}{Fig}{Figs}
\crefname{table}{Tab}{Tabs}
\crefname{equation}{eq.}{eqs.}

\leadauthor{Siddhad} % Please give the surname of the lead author for the running footer

\begin{document}

\title{MECASA: Motor Execution Classification using Additive Self-Attention for Hybrid EEG-fNIRS Data}
\shorttitle{MECASA: Motor Execution Classification using Additive Self-Attention for Hybrid EEG-fNIRS Data}

\author[1,\Letter]{Gourav~Siddhad~\orcidlink{0000-0001-5883-3863}}
\author[2]{Juhi~Singh~\orcidlink{0009-0003-1322-8864}}
\author[1]{Partha~Pratim~Roy~\orcidlink{0000-0002-5735-5254}}
\affil[1]{Department of Computer Science and Engineering, Indian Institute of Technology, Roorkee, Uttarakhand, 247667, India}
\affil[2]{Department of Artificial Intelligence \& Machine Learning, Manipal University, Jaipur, Rajasthan, 303007, India}

\maketitle

%%%%%%%%%%%%%%%%%%%%%%%%%%%%%%%%%%%%%%%%%%%%%%%%%%
%%%%%%%%%%%%%%%%%%%%%%%%%%%%%%%%%%%%%%%%%%%%%%%%%%

\begin{abstract}
Motor execution, a fundamental aspect of human behavior, has been extensively studied using BCI technologies. EEG and fNIRS have been utilized to provide valuable insights, but their individual limitations have hindered performance. This study investigates the effectiveness of fusing electroencephalography (EEG) and functional near-infrared spectroscopy (fNIRS) data for classifying rest versus task states in a motor execution paradigm. Using the SMR Hybrid BCI dataset, this work compares unimodal (EEG and fNIRS) classifiers with a multimodal fusion approach. It proposes Motor Execution using Convolutional Additive Self-Attention Mechanisms (MECASA), a novel architecture leveraging convolutional operations and self-attention to capture complex patterns in multimodal data. MECASA, built upon the CAS-ViT architecture, employs a computationally efficient, convolutional-based self-attention module (CASA), a hybrid block design, and a dedicated fusion network to combine features from separate EEG and fNIRS processing streams. Experimental results demonstrate that MECASA consistently outperforms established methods across all modalities (EEG, fNIRS, and fused), with fusion consistently improving accuracy compared to single-modality approaches. fNIRS generally achieved higher accuracy than EEG alone. Ablation studies revealed optimal configurations for MECASA, with embedding dimensions of 64-128 providing the best performance for EEG data and OD128 (upsampled optical density) yielding superior results for fNIRS data. This work highlights the potential of deep learning, specifically MECASA, to enhance EEG-fNIRS fusion for BCI applications.
\end{abstract}

\begin{corrauthor}
g\_siddhad\at cs.iitr.ac.in
\end{corrauthor}

\section*{Impact Statement}
This research presents a novel deep learning architecture, MECASA, for enhanced classification of motor execution tasks using combined EEG and fNIRS data. By effectively fusing these complementary modalities, MECASA achieves significant performance improvements compared to existing methods. This advancement has important implications for Brain-Computer Interface (BCI) technology, particularly in the fields of motor rehabilitation and cognitive assessment. The improved accuracy and robustness of MECASA could lead to more effective and reliable BCI-based interventions for individuals with motor impairments, enabling more precise control of assistive devices, improved neurofeedback training, and more accurate assessment of cognitive states related to motor function. The computationally efficient CASA module also paves the way for potential real-time applications of the proposed method, which is crucial for practical BCI systems. Furthermore, the findings regarding optimal data representations (embedding dimensions for EEG and OD128 for fNIRS) provide valuable insights for future research in multimodal BCI signal processing.

%%%%%%%%%%%%%%%%%%%%%%%%%%%%%%%%%%%%%%%%%%%%%%%%%%
%%%%%%%%%%%%%%%%%%%%%%%%%%%%%%%%%%%%%%%%%%%%%%%%%%

\section{Introduction}
\label{sec_intro}

Motor execution, the overt and volitional movement central to human activity, remains a critical focus in neuroscience research. Understanding the neural mechanisms underpinning motor execution is vital for mapping brain regions such as the motor cortex and developing effective rehabilitation strategies~\cite{brigadoi2012exploring}. Brain-computer interfaces (BCIs) have emerged as powerful tools in this domain, enabling researchers to explore motor execution through neural signals. Among the most widely used modalities for BCI applications are Electroencephalography (EEG) and Functional Near-Infrared Spectroscopy (fNIRS), each providing distinct insights into brain activity.

EEG, a non-invasive technique with high temporal resolution, has long been a cornerstone of cognitive neuroscience research~\cite{morrone2024eeg}. Despite its ability to capture rapid neural dynamics, EEG's low spatial resolution limits its capacity to precisely localize brain activity. Conversely, fNIRS provides a detailed view of cortical hemodynamics by measuring changes in oxygenated and deoxygenated hemoglobin concentrations. While fNIRS suffers from the delayed nature of hemodynamic responses, it is less susceptible to motion artifacts and remains effective in motor paradigms~\cite{brigadoi2012exploring}.

Recent advancements in deep learning have further enhanced the ability to extract relevant features from these data and classify motor-related tasks~\cite{ma2017extraction}. For instance, EEG oscillations in the 4–12Hz range have been shown to modulate during both motor execution and observation~\cite{frenkel2013dynamics}. Meanwhile, deep learning models applied to fNIRS data have achieved success in tasks like cognitive state classification and driver drowsiness detection~\cite{chaisaen2020decoding}. Deep learning approaches have also demonstrated the potential to identify multisensory features in motor execution tasks from EEG data~\cite{elsayed2021deep}.

fNIRS has gained traction for its utility in motor research, particularly in investigating motor imagery and execution~\cite{coyle2004suitability}. Compared to more traditional methods like functional Magnetic Resonance Imaging (fMRI), fNIRS offers significant advantages, such as portability and reduced sensitivity to motion artifacts~\cite{koehler2012human}. Studies have shown that distinct hemodynamic patterns associated with motor imagery can be captured effectively with fNIRS, and deep learning models have been successfully applied to classify cognitive states and detect driver drowsiness using fNIRS data~\cite{karmakar2023real}.

Despite the strengths of both EEG and fNIRS, each modality comes with limitations. EEG's low spatial resolution can hinder precise localization of brain activity, while fNIRS, with its reliance on hemodynamic signals, is less capable of capturing rapid neural events. To address these limitations, multimodal data fusion techniques have been introduced, combining EEG and fNIRS to offer a more comprehensive understanding of cognitive functions. Research has demonstrated that such fusion can enhance BCI performance and improve the detection of mental states like stress~\cite{tanveer2019enhanced}. Fusion strategies in multimodal brain-computer interfaces (BCIs) typically involve combining data from different modalities at various stages of processing. While early-stage fusion has shown promising results, a variety of techniques have been explored to optimize the integration of electroencephalography (EEG) and functional near-infrared spectroscopy (fNIRS) data.

In motor paradigms, EEG-fNIRS fusion has significantly enhanced BCI performance. For instance, studies have demonstrated improved classification accuracy for motor imagery tasks involving imagined hand clenching and ankle joint movements~\cite{yin2015hybrid,al2021bimodal}. These findings highlight the potential of multimodal approaches for advancing BCIs in applications such as lower limb rehabilitation. By combining EEG's temporal resolution with fNIRS's spatial resolution, EEG-fNIRS fusion can address the limitations of each modality and provide a more comprehensive understanding of brain activity. However, the optimal fusion strategy for specific applications remains an area of ongoing research. Future work should explore novel fusion techniques to unlock the full potential of EEG-fNIRS integration in motor paradigms.

In this study, the effectiveness of EEG-fNIRS fusion in classifying rest versus task in a motor execution paradigm is investigated. Using the SMR Hybrid BCI dataset, unimodal EEG and fNIRS classifiers with a multimodal fusion approach are compared, extracting feature vectors from each modality independently before applying a classifier to perform the classification task. Furthermore, Motor Execution using Convolutional Additive Self-Attention Mechanisms (MECASA) is proposed, which outperformed all other models tested in this study. MECASA leverages the power of convolutional operations and self-attention mechanisms to effectively capture complex patterns in multimodal data, demonstrating the potential of deep learning in enhancing EEG-fNIRS fusion for BCI applications.

This paper introduces MECASA (Motor Execution Convolutional Additive Self-Attention), a novel architecture for classifying motor execution tasks using fused EEG and fNIRS data. Key contributions include:
\begin{itemize}
    \item A computationally efficient, convolutional-based self-attention module (CASA) replaces traditional self-attention ($O(N)$ complexity).
    \item A hybrid block design (with an Integration subnet, CASA module, and MLP with residual shortcuts) and a dedicated fusion network combine features from separate EEG and fNIRS processing streams, leveraging their complementary information (EEG's high temporal resolution and fNIRS's spatial/hemodynamic information).
    \item MECASA consistently outperformed established methods across all modalities (EEG, fNIRS, and fused), with fusion consistently improving accuracy compared to single-modality approaches, and fNIRS generally achieving higher accuracy than EEG alone.
\end{itemize}

The research paper is organized into five key sections to provide a comprehensive analysis of the proposed work. Section~\ref{sec_related} presents a detailed literature review, highlighting the existing methods and approaches related to the research problem, while identifying gaps that this study aims to address. Section~\ref{sec_method} outlines the methodology, with a focus on the architectural design and implementation of the MECASA framework, explaining its components and their functionality. Section~\ref{sec_results} is dedicated to the results and discussion, starting with a description of the dataset used, followed by implementation details and a comparative analysis of classifiers to evaluate the performance of the proposed model. This section also includes an ablation study to assess the contribution of individual components of the framework. Finally, Section~\ref{sec_conclusion} concludes the paper by summarizing the findings, discussing the implications of the results, and suggesting potential directions for future research.

%%%%%%%%%%%%%%%%%%%%%%%%%%%%%%%%%%%%%%%%%%%%%%%%%%
%%%%%%%%%%%%%%%%%%%%%%%%%%%%%%%%%%%%%%%%%%%%%%%%%%

\section{Related Work}
\label{sec_related}

This section reviews motor execution (ME) classification in brain-computer interfaces (BCIs), covering its significance, challenges, and employed techniques.

\subsection{Motor Execution Classification and its Challenges}

ME classification is crucial for neuroscience and BCI development. In neuroscience, it enhances understanding of brain movement control and impacts neurorehabilitation for motor disabilities like multiple sclerosis and hemiplegia~\cite{asanza2022identification}. Accurate motor task classification improves assistive technologies, enhancing quality of life. In BCIs, accurate ME signal classification is essential for effective brain-device communication, offering independence to individuals with lost mobility through prosthetic limb or exoskeleton control~\cite{hamid2022analyzing}. EEG and other neuroimaging techniques like fNIRS have shown promise~\cite{shuqfa2023decoding}. Recent studies explore deep learning (CNNs, DBNs) for automatic feature extraction and classification from EEG/fNIRS signals, achieving higher accuracy than traditional methods (SVM, LDA)~\cite{hamid2022analyzing}. Novel algorithms like Riemannian geometry decoding and DMNN address EEG signal non-stationarity and noise~\cite{shuqfa2023decoding}.

However, EEG motor execution classification faces challenges impacting BCI accuracy and efficiency, including signal noise and low SNR, high inter-subject variability, signal non-stationarity, and complexities in multimodal data fusion. Noise complicates meaningful information extraction, mitigated by techniques like sparse spectrotemporal decomposition and multiscale principal component analysis~\cite{sun2020eeg}. Inter-subject variability hinders generalized models; transfer learning and domain adaptation address this by aligning data from different subjects~\cite{chu2024transfer}. The non-stationary nature of EEG signals requires adaptive models; self-attention mechanisms and adaptive feature extraction have shown promise~\cite{yang2024diagonal}. Integrating EEG with other modalities like MEG offers improved performance through multimodal data fusion~\cite{wang2024msfnet}. Advanced deep learning models, particularly multi-layer CNNs and attention-based models, capture complex spatio-temporal EEG features~\cite{yang2024diagonal}. Feature fusion techniques, such as multi-scale space-time frequency fusion and diagonal masking self-attention networks, enhance feature extraction and integration~\cite{wang2024msfnet}. Optimizing EEG channel selection can reduce noise and computational burden~\cite{khabti2024optimal}. Future research will likely focus on improving model generalizability and exploring new signal processing techniques for enhanced real-world BCI performance~\cite{wang2022brain}.

\subsection{Classification Algorithms for Motor Execution}

Traditional machine learning algorithms like SVMs and Random Forests have been widely used. SVMs, effective with high-dimensional data, achieve reasonable accuracy, though they may not always outperform deep learning~\cite{batistic2023motor}. Combining SVMs with other techniques can enhance performance, but careful hyperparameter tuning and feature selection are needed~\cite{hou2022novel}. Random Forests, robust to noise, effectively distinguish between real and imagery motor activities~\cite{de2023electroencephalography}. However, intersubject EEG variability can impact their accuracy. While deep learning often achieves higher accuracy, SVMs and Random Forests remain competitive due to their simplicity and lower computational demands~\cite{batistic2023motor}. A key challenge is their reliance on potentially complex feature extraction and selection~\cite{de2023electroencephalography}.

Deep learning models have become increasingly prominent. CNNs are widely used due to their ability to extract spatio-temporal features directly from raw EEG data~\cite{amin2019deep}. Hybrid CNN architectures (CNNs with LSTM/BiLSTM) further enhance accuracy~\cite{boutarfaia2023deep}. Hybrid models combining CNNs with autoencoders like VAEs have also shown improved performance~\cite{dai2019eeg}. Hybrid-scale CNNs address variability in optimal convolution scales across subjects~\cite{dai2020hs}. Transfer learning has shown significant improvements in handling limited subject-specific data~\cite{zhang2021adaptive}. Challenges remain, including overfitting, which can be mitigated by data augmentation and hyperparameter optimization, and computational complexity~\cite{al2021deep}.

\subsection{Multimodal Approaches: EEG, fNIRS, and their Fusion}

EEG-only approaches are limited by low spatial resolution, hindering accurate classification of motor imagery tasks, especially distinguishing between closely related movements~\cite{hu2023cross}. This is compounded by many algorithms operating within a single spatial domain. EEG signals also suffer from poor SNR and consistency, exacerbated by signal distortions and intersubject variability. Data acquisition and limited training data also present challenges~\cite{sartipi2024subject}.

fNIRS offers a complementary perspective by measuring hemodynamic responses~\cite{khani2023fusion}. By capturing the spatial dynamics of blood oxygenation, fNIRS reveals hemispheric lateralization during motor tasks~\cite{hramov2020functional}. Studies have demonstrated high classification accuracy for real motor movements using fNIRS~\cite{hramov2020functional}. Integrating deep learning techniques, such as CNNs, has significantly enhanced fNIRS signal classification accuracy~\cite{khani2023fusion}. Advancements focus on developing subject-independent models~\cite{erdogan2019classification}. However, fNIRS also presents challenges, requiring effective signal processing techniques to mitigate noise and aliasing~\cite{peng2018single}. Developing real-time classification systems is also a key research area~\cite{hosni2020fnirs}.

Hybrid EEG-fNIRS systems leverage the strengths of both modalities, providing a more comprehensive view of brain activity~\cite{croce2017exploiting}. This synergistic approach has demonstrated improved motor task decoding, particularly for MI tasks~\cite{xu2023motor}. Methodological approaches include feature extraction and processing techniques like CSP and PCA, and deep learning frameworks like FGANet~\cite{kwak2022fganet}. Hybrid systems have been applied to various motor execution tasks~\cite{khani2023fusion}. However, challenges remain in integrating and synchronizing the two modalities due to their differing temporal resolutions. Early fusion methods are being explored to address this~\cite{kwak2022fganet}. Future research will likely focus on optimizing feature extraction algorithms and improving the physical design of hybrid systems~\cite{li2024hybrid}.

\subsection{Attention Mechanisms in BCIs}

Attention mechanisms have become essential in neural networks, improving performance by focusing on relevant features. Integrating attention improves both accuracy and efficiency~\cite{jiang2024generalised}. Attention mechanisms are particularly relevant for complex data like EEG and hybrid EEG-fNIRS. In EEG-based models, channel-wise attention improves discriminative feature extraction and hybrid attention networks leverage spatial-temporal correlations~\cite{zhao2023hybrid}. For motor imagery EEG decoding, integrating attention into multi-scale fusion CNNs enhances classification~\cite{li2020multi}. In hybrid EEG-fNIRS models, attention mechanisms in early fusion structures (e.g., FGANet) spatially align signals~\cite{kwak2022fganet}.

Despite these benefits, challenges remain, including understanding and interpreting attention weights and cross-patient generalization~\cite{zhao2023hybrid}. Traditional attention mechanisms also struggle with multi-scale information fusion~\cite{tao2023adfcnn}. In motor execution classification, integrating attention mechanisms has shown significant promise. The ADFCNN model demonstrates improved motor imagery classification by effectively fusing spectral and spatial EEG information~\cite{tao2023adfcnn}. Spiking neural networks (SNNs) with attention, like TCJA-SNN, offer potential for high-performance, energy-efficient computing~\cite{zhu2024tcja}. However, balancing performance gains with energy efficiency in SNNs remains a key challenge~\cite{zhu2024tcja}. Moreover, the application of attention mechanisms in this domain is still relatively nascent, necessitating future research to prioritize generalization across diverse domains and tasks~\cite{niu2021review}.

Leveraging additive self-attention addresses key challenges in EEG-based motor imagery (MI) classification, such as non-stationarity and low signal-to-noise ratio, by enhancing feature extraction and fusion~\cite{yang2024diagonal}. It allows for better feature emphasis and extraction, facilitating the analysis of global information. Self-attention enables improved information fusion from different scales, capturing both spectral and spatial information~\cite{tao2023adfcnn}. It also enhances model generalization across datasets and conditions. Models incorporating self-attention, such as DMSA-MSNet, have demonstrated state-of-the-art performance~\cite{yang2024diagonal}.

This review has considered hybrid EEG-fNIRS systems and the application of attention mechanisms, including additive self-attention, in BCIs. While hybrid systems offer improved spatiotemporal information and attention mechanisms enhance model performance, challenges related to integration, interpretability, computational cost, and cross-subject generalization remain. Future research should prioritize addressing these issues.

%%%%%%%%%%%%%%%%%%%%%%%%%%%%%%%%%%%%%%%%%%%%%%%%%%
%%%%%%%%%%%%%%%%%%%%%%%%%%%%%%%%%%%%%%%%%%%%%%%%%%

\begin{figure*}[!t]
    \centering
    \includegraphics[width=\linewidth]{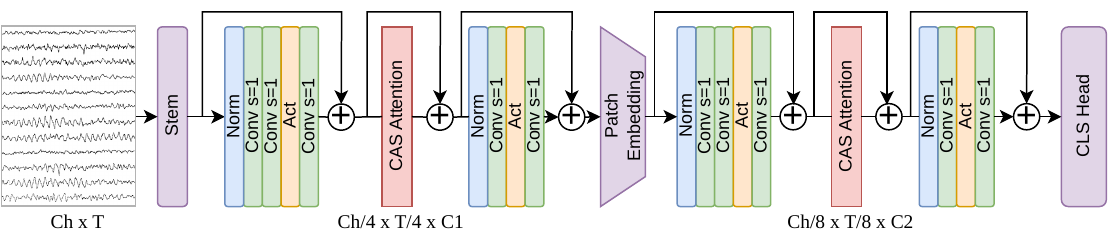}
    \caption{Illustration of the proposed classification backbone. Two stages downsample the original signal.}
    \label{fig_method_arch}
\end{figure*}

\begin{figure}[!t]
    \centering
    \includegraphics[width=0.9\linewidth]{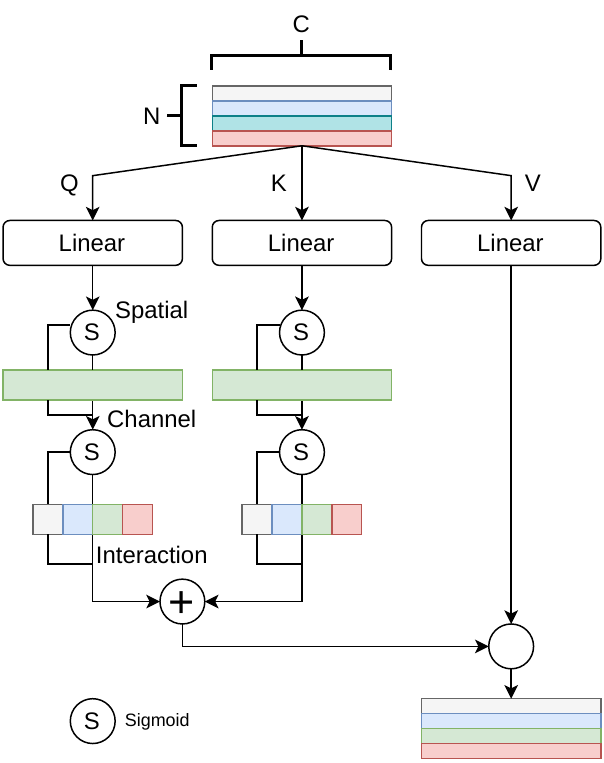}
    \caption{Convolution Adaptive Separable Attention}
    \label{fig_method_casa}
\end{figure}

\begin{figure}[!t]
    \centering
    \includegraphics[width=0.4\linewidth]{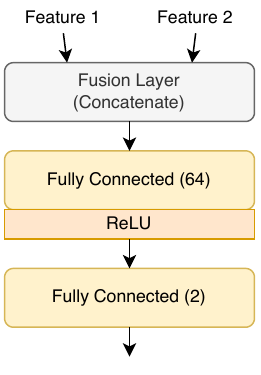}
    \caption{Architecture of the proposed fusion network used for classification of combined features of EEG and fNIRS data}
    \label{fig_method_fusion}
\end{figure}

\section{Methodology}
\label{sec_method}

Building upon the success of the Convolution Additive Self-Attention Vision Transformer (CAS-ViT) architecture~\cite{zhang2024cas}, this study proposes an approach for classifying Motor Execution tasks using EEG and fNIRS data. The core component of this methodology is the CASA module, which effectively combines convolutional operations and self-attention to capture both local and global dependencies within the neural signals. This section will delve into the details of the attention mechanism and the proposed architectural design (shown in Figure~\ref{fig_method_arch}).

\subsection{Self-Attention}

The self-attention mechanism is a key component of transformer architectures that allows models to capture relationships between different elements within a sequence. It works by calculating a weighted sum of input elements, where the weights are determined based on the similarity between the current element and all other elements. Mathematically, the self-attention mechanism can be expressed as:
\begin{equation}
    O_i = \sum^N_{j=1} \frac{Sim \left( Q_i, K_j \right)}{\sum^N_{j=1} Sim \left( Q_i, K_j \right)} V_j \enspace,
    \label{eq:self-attention-vector}
\end{equation}
where $Sim(Q_i, K_j)$ is a similarity function (e.g., dot product, cosine similarity) between the query $Q_i$ and key $K_j$, and $V_j$ is the value associated with the j-th element.

Classical transformer structure employs $Sim(Q, K) = \text{exp} \left( Q K^\top / \sqrt{d} \right)$, in this case the self-attention mechanism can be represented by $Softmax(\cdot)$ function:
\begin{equation}
    O = Softmax \left( \frac{Q K^\top}{\sqrt{d}} \right) V \enspace.
    \label{eq:self-attention-softmax}
\end{equation}

\subsection{Convolutional Additive Self-attention}
\label{subsection:cas}

Traditional self-attention mechanisms often involve complex matrix operations, which can be computationally expensive and Softmax often hinders efficient inference, limiting their practical applicability. To address this, Convolutional Additive Self-Attention (CASA) is employed in this study. CASA leverages convolutional operations to efficiently compute similarities between elements, making it more suitable for real-time applications. The core idea of CASA is to replace the traditional similarity function with a convolutional-based one. This allows the model to capture both local and global dependencies within the input sequence. As depicted in Figure~\ref{fig_method_casa}, the similarity between $Q \in \mathbb{R}^{N \times d}$ and $K \in \mathbb{R}^{N \times d}$ is calculated as:
\begin{equation}
    Sim \left( Q,K \right) = \Phi (Q) + \Phi(K) \quad \text{where} \quad \Phi(Q) = \mathcal{C}(\mathcal{S} (Q)) \enspace,
    \label{eq:sim-additive}
\end{equation}
Here, Query, Key, and Value are obtained through independent linear transformations, such as $Q = W_q x$, $K = W_k x$, $V = W_v x$. The context mapping function $\Phi(\cdot)$ encapsulates the essential information interactions and can be implemented using convolutional operations, offering flexibility in design.

$\Phi(\cdot)$ can be concretize as a combination of Sigmoid-based channel attention $\mathcal{C}(\cdot) \in \mathbb{R}^{N \times d}$ and spatial attention $\mathcal{S}(\cdot) \in \mathbb{R}^{N \times d}$. The output of CASA mechanism is then given by:
\begin{equation}
    O = \Gamma \left( \Phi (Q) + \Phi(K) \right) \cdot V \enspace,
    \label{eq:self-attention-cas}
\end{equation}
where $\Gamma(\cdot) \in \mathbb{R}^{N \times d}$ is a linear transformation for integrating the contextual information. The convolutional nature of CASA operations results in a complexity of $\mathcal{O}(N)$, making it computationally efficient.

\subsection{MECASA Architecture}

Figure~\ref{fig_method_arch} illustrates the proposed network architecture of MECASA. The input signal, of size $Ch \times T$, is downsampled to $\frac{Ch}{4} \times \frac{T}{4} \times C_1$ using two consecutive convolutional layers with stride 2 in the Stem. Next, the downsampled features pass through four stage encoding layers. Between each stage, Patch Embedding layers are employed to further downsample by a factor of 2, resulting in feature maps of size $\frac{Ch}{8} \times \frac{T}{8} \times C_2$. The index $i \in \{1,2\}$ denotes the feature map channels, which remains constant throughout the blocks.

The block design draws inspiration from hybrid networks such as EfficientViT~\cite{liu2023efficientvit} and EdgeViT~\cite{pan2022edgevits} and comprises three primary components with residual shortcuts: the Integration subnet, CASA, and MLP. The Integration subnet, inspired by SwiftFormer~\cite{shaker2023swiftformer}, consists of three depth-wise convolutional layers activated by ReLU~\cite{glorot2011deep}.

\subsection{Fusion Network}

The fusion network, designed to classify EEG and fNIRS features (shown in Fig.~\ref{fig_method_fusion}), integrates the strengths of these two modalities to enhance classification accuracy. The features (feature 1 and feature 2) used in the fusion network for classification are the features extracted from the last layers of their respective networks before the classification heads.

The process begins with feature extraction, where separate, modality-specific networks are used to process data from EEG and fNIRS. The EEG-specific network is designed to capture the temporal and spectral characteristics of electrical brain activity. Concurrently, the fNIRS-specific network extracts features that reflect hemodynamic responses, such as changes in blood oxygenation.

Once these features are extracted by their respective networks, they are combined in a fusion layer through concatenation. This step integrates the complementary information provided by the two modalities, enabling the network to learn a richer representation of the data. The fusion layer serves as a bridge, combining the high temporal resolution of EEG with the spatial and hemodynamic information of fNIRS.

The fused features are then passed through a series of fully connected layers. A first fully connected layer, typically with a moderate number of neurons (e.g., 64), applies a ReLU activation function. This introduces non-linearity and allows the network to capture complex relationships in the data. Following this, a second fully connected layer, with a smaller number of neurons corresponding to the output classes (e.g., 2 neurons for binary classification), generates the final classification output.

The final output layer uses an appropriate activation function, such as softmax for multi-class classification or sigmoid for binary classification, to produce class probabilities. This hybrid EEG-fNIRS approach leverages the complementary nature of the two modalities, combining EEG’s high temporal resolution and fNIRS’s hemodynamic information to improve classification performance. By addressing the limitations of single-modality systems, this fusion network enhances robustness and accuracy in diverse classification tasks.

%%%%%%%%%%%%%%%%%%%%%%%%%%%%%%%%%%%%%%%%%%%%%%%%%%
%%%%%%%%%%%%%%%%%%%%%%%%%%%%%%%%%%%%%%%%%%%%%%%%%%

\section{Results and Discussion}
\label{sec_results}

\subsection{Experimental Data}
The SMR Hybrid EEG-fNIRS dataset~\cite{buccino2016hybrid} was used, comprising 15 healthy, right-handed male participants (mean age: 27.4 ± 7.7 years). Participants performed five blocks of motor execution tasks, each with 20 randomized trials (5 trials per movement). EEG and fNIRS data were collected during 12-second trials, including a 6-second rest period and a 6-second movement phase. A NIRScout 8–16 fNIRS system with 34 channels and a microEEG system with 21 channels were used to record cortical activity and EEG signals, respectively. Both systems were mounted on an extended EEG cap.

To prepare EEG and fNIRS data for the classification of cognitive states (Rest vs. Task), pre-processing stage was implemented to minimize artifacts and enhance computational efficiency. EEG signals were first band-pass filtered between 0.5-45 Hz and then downsampled to 128 Hz. For fNIRS data processing, the raw data was converted into the optical density (OD) format before being downsampled to a frequency of 10 Hz (OD10). Subsequently, it was upsampled to 128 Hz (OD128). The Modified Beer-Lambert law~\cite{kocsis2006modified} was applied, then converted to optical density and then upsampled to 128 Hz. These changes were combined to yield total hemoglobin (HbT). Following these steps, both EEG and fNIRS data were segmented into one-second epochs with a 0.5-second overlap, resulting in standardized dimensions of (1, 21, 128) and (1, 68, 128) for EEG and fNIRS (OD128), respectively, per epoch. The entire dataset, comprising 41,228 samples for each modality, was split into training (70\%), validation (15\%), and test sets (15\%) for model development and evaluation.

%%%%%%%%%%%%%%%%%%%%%%%%%%%%%%%%%%%%%%%%%%%%%%%%%%
%%%%%%%%%%%%%%%%%%%%%%%%%%%%%%%%%%%%%%%%%%%%%%%%%%

\subsection{Implementation Details}
The computational environment consisted of an AMD Ryzen 7 2700X CPU and an NVIDIA RTX A4000 16GB GPU running Ubuntu 20.04. This hardware configuration facilitated the implementation of Deep Learning (DL) models using Python 3.12 and the PyTorch library. The Adam optimizer, known for its efficiency, was used with its default hyperparameters ($\eta$=0.001, $\beta_1$=0.9, $\beta_2$=0.999). All DL models were trained for 100 epochs, using a batch size of 16 and a learning rate of 1e-4. Stratified five-fold cross-validation was employed to evaluate classification accuracy, with the results averaged for a comprehensive assessment.

%%%%%%%%%%%%%%%%%%%%%%%%%%%%%%%%%%%%%%%%%%%%%%%%%%
%%%%%%%%%%%%%%%%%%%%%%%%%%%%%%%%%%%%%%%%%%%%%%%%%%

\subsection{Classifiers}
This work employs a balanced evaluation approach using three established classifiers for EEG-based emotion classification. EEGNet~\cite{lawhern2018eegnet} is a CNN-based architecture that achieves competitive accuracy using deep and separable convolutions. It incorporates temporal convolution for learning frequency filters, depth-wise convolution for frequency-specific spatial filters, and separable convolution for efficient feature map combinations. TSception~\cite{ding2022tsception} utilizes a dynamic temporal layer to learn temporal and frequency representations from EEG channels. It also includes an asymmetric spatial layer for capturing global spatial patterns and emotional asymmetry, a high-level fusion layer, and a final classifier that leverages various convolutional kernel sizes for spatial analysis. ConvNext~\cite{liu2022convnet,siddhad2024neural} is a state-of-the-art CNN architecture that achieves competitive performance on various image classification benchmarks. It incorporates design principles from recent transformer models to enhance feature learning and improve efficiency compared to traditional CNNs. LMDA-Net~\cite{miao2023lmda} is a lightweight deep-learning model specifically designed for EEG-based emotion classification. It employs a multi-modal approach, combining temporal and spatial features, to effectively capture the complex patterns in EEG signals, resulting in efficient and accurate emotion recognition. Transformer~\cite{siddhad2024efficacy} networks, initially designed for natural language processing tasks, have gained significant attention for their potential in modeling sequential data. Their unique architecture, characterized by attention mechanisms and self-attention layers, enables them to capture long-range dependencies and parallel processing, making them well-suited for analyzing EEG signals. These complex biological signals exhibit intricate temporal and spatial patterns, which can be effectively modeled by the transformer's ability to learn global relationships and dependencies within the data.

%%%%%%%%%%%%%%%%%%%%%%%%%%%%%%%%%%%%%%%%%%%%%%%%%%
%%%%%%%%%%%%%%%%%%%%%%%%%%%%%%%%%%%%%%%%%%%%%%%%%%

\subsection{Evaluation}
\begin{table}[!t]
    \centering
    \caption{Result for EEG and fNIRS data for classification of Rest vs Task, accuracy with 95\% confidence interval}
    \resizebox{\linewidth}{!}{
    \begin{tabular}{l c c c}
        \toprule
        \textbf{Method} & \textbf{EEG} & \textbf{fNIRS} & \textbf{Fusion} \\
        \midrule
        EEGNet~\cite{lawhern2018eegnet} & $72.40 \pm 0.52$ & $74.80 \pm 0.50$ & $83.38 \pm 0.40$ \\
        TSception~\cite{ding2022tsception} & $72.07 \pm 0.70$ & $81.45 \pm 1.18$ & $82.93 \pm 0.80$\\
        Transformer~\cite{siddhad2024efficacy} & $60.25 \pm 0.53$ & $66.47 \pm 1.07$ & $67.09 \pm 1.38$ \\
        LMDA~\cite{miao2023lmda} & $69.60 \pm 0.80$ & $74.47 \pm 0.54$ & $57.47 \pm 0.50$ \\
        ConvNeXT~\cite{siddhad2024neural} & $67.23 \pm 0.25$ & $78.43 \pm 4.65$ & $80.17 \pm 5.38$ \\
        \midrule
        \textbf{MECASA} & $\mathbf{75.07 \pm 3.89}$ & $\mathbf{86.52 \pm 1.38}$ & $\mathbf{87.34 \pm 0.42}$\\
        \bottomrule
    \end{tabular}}
    \label{tab_result}
\end{table}

Table~\ref{tab_result} summarizes the performance of various deep learning methods on EEG and fNIRS data for the `Rest vs. Task' classification task. The accuracy of each method is reported with a 95\% confidence interval. The results demonstrate that MECASA consistently outperforms all other methods, achieving the highest accuracy across EEG, fNIRS, and fused modalities. This finding highlights MECASA's effectiveness in leveraging the complementary information from both EEG and fNIRS to accurately distinguish between `Rest' and `Task' states. Furthermore, it is observed that fNIRS data consistently yielded higher classification accuracies than EEG data, suggesting its potential as a more informative modality for this specific task. This could be attributed to fNIRS's ability to directly measure changes in brain hemodynamics, which are closely linked to cognitive processes. The fusion of EEG and fNIRS data generally resulted in improved accuracy compared to using either modality alone, emphasizing the benefits of multi-modality approaches. This suggests that combining the temporal resolution of EEG with the metabolic information from fNIRS can provide a more comprehensive understanding of brain activity during `Rest' and `Task' states. In addition to MECASA, EEGNet, TSception, and ConvNeXT also demonstrated promising performance, indicating their suitability for EEG-fNIRS based classification tasks.

%%%%%%%%%%%%%%%%%%%%%%%%%%%%%%%%%%%%%%%%%%%%%%%%%%
%%%%%%%%%%%%%%%%%%%%%%%%%%%%%%%%%%%%%%%%%%%%%%%%%%

\subsection{Ablation Study}

\begin{table}[!t]
    \centering
    \caption{Ablation for CASA for EEG data for classification of Rest vs Task, accuracy with 95\% confidence interval}
    \begin{tabular}{l c c}
        \toprule
        \textbf{Embedding Dims} & \textbf{EEG} & \textbf{fNIRS} \\
        \midrule
        16-32 & $71.85 \pm 2.08$ & $85.37 \pm 0.50$ \\
        32-64 & $73.21 \pm 1.43$ & $86.20 \pm 0.74$ \\
        48-56 & $72.81 \pm 5.24$ & $87.57 \pm 0.69$ \\
        64-128 & $75.07 \pm 3.89$ & $86.52 \pm 1.38$ \\
        \bottomrule
    \end{tabular}
    \label{tab_ablation_casa}
\end{table}

\begin{table}[!t]
    \centering
    \caption{Ablation for fNIRS data for classification of Rest vs Task, accuracy with 95\% confidence interval}
    \resizebox{\linewidth}{!}{
    \begin{tabular}{l c c c}
        \toprule
        \textbf{Method} & \textbf{OD10} & \textbf{HBT} & \textbf{OD128} \\
        \midrule
        EEGNet~\cite{lawhern2018eegnet} & $74.59 \pm 0.66$ & $68.62 \pm 0.48$ & $74.80 \pm 0.50$ \\
        TSception~\cite{ding2022tsception} & $72.90 \pm 1.76$ & $73.28 \pm 1.06$ & $81.45 \pm 1.18$ \\
        LMDA~\cite{miao2023lmda} & $72.69 \pm 0.47$ & $65.72 \pm 0.75$ & $74.47 \pm 0.54$ \\
        Transformer~\cite{siddhad2024efficacy} & $62.13 \pm 0.81$ & $65.49 \pm 1.90$ & $66.47 \pm 1.07$ \\
        ConvNeXT~\cite{siddhad2024neural} & $75.85 \pm 1.49$ & $77.02 \pm 2.12$ & $78.43 \pm 4.65$ \\
        \midrule
        \textbf{MECASA} & $\mathbf{77.51 \pm 1.30}$ & $\mathbf{80.06 \pm 1.55}$ & $\mathbf{86.52 \pm 1.38}$\\
        \bottomrule
    \end{tabular}}
    \label{tab_ablation_fnirs}
\end{table}

Table~\ref{tab_ablation_casa} presents an ablation study to evaluate the impact of embedding dimensions on the CASA model's performance for EEG data. The results indicate that the choice of embedding dimensions is a critical factor in classification accuracy. Embeddings of 64-128 dimensions consistently outperform other configurations, suggesting that this range strikes a balance between capturing sufficient information and avoiding overfitting. Table~\ref{tab_ablation_fnirs} further explores the influence of fNIRS data representation on the performance of various methods. The study compares three representations: OD10, HbT, and OD128. The findings demonstrate that OD128, which offers a higher sampling frequency, generally leads to superior classification accuracy compared to OD10 and HbT. This suggests that capturing more detailed temporal information from fNIRS data is beneficial for improving model performance.

MECASA emerges as a consistently top-performing method across all fNIRS data representations. This highlights its robustness and effectiveness in leveraging the unique characteristics of fNIRS data. Other methods exhibit varying sensitivities to the choice of fNIRS representation, indicating that their performance may be influenced by specific data characteristics. In conclusion, these ablation studies offer valuable insights into the factors influencing the CASA model's performance. The choice of embedding dimensions for EEG data and the selection of fNIRS data representation are both crucial considerations for achieving optimal classification results. The consistent superiority of MECASA across different data modalities underscores its potential as a promising approach for EEG-fNIRS based classification tasks.

%%%%%%%%%%%%%%%%%%%%%%%%%%%%%%%%%%%%%%%%%%%%%%%%%%
%%%%%%%%%%%%%%%%%%%%%%%%%%%%%%%%%%%%%%%%%%%%%%%%%%

\section{Conclusion}
\label{sec_conclusion}

This study demonstrates the efficacy of combining EEG and fNIRS data for motor execution classification tasks. Multimodal fusion consistently outperformed unimodal approaches, emphasizing the benefits of leveraging complementary information from both modalities. The MECASA model, incorporating convolutional additive self-attention, emerged as a superior deep learning architecture, effectively capturing and integrating the unique characteristics of EEG and fNIRS data. Ablation studies highlighted the significance of selecting optimal embedding dimensions for EEG and data representations for fNIRS. These findings underscore the importance of careful feature engineering in BCI applications. The MECASA model's strong performance suggests its potential for enhancing BCI systems, particularly in motor rehabilitation and stress detection. Future research should explore refining fusion strategies and investigating the broader applicability of the MECASA model to other cognitive tasks. Addressing the challenges associated with real-world BCI deployment is also a key direction for future work. By addressing these areas, the field can further advance the development of effective and reliable BCI technologies.

%%%%%%%%%%%%%%%%%%%%%%%%%%%%%%%%%%%%%%%%%%%%%%%%%%
%%%%%%%%%%%%%%%%%%%%%%%%%%%%%%%%%%%%%%%%%%%%%%%%%%

\section*{Bibliography}
\bibliography{references}

%%%%%%%%%%%%%%%%%%%%%%%%%%%%%%%%%%%%%%%%%%%%%%%%%%
%%%%%%%%%%%%%%%%%%%%%%%%%%%%%%%%%%%%%%%%%%%%%%%%%%

\end{document}